\documentclass[sigconf]{acmart}
\usepackage{multirow}
\usepackage[normalem]{ulem}
\usepackage{threeparttable}
\useunder{\uline}{\ul}{}
\usepackage{subcaption}
\usepackage{pifont}
\AtBeginDocument{%
  }

\renewcommand\footnotetextcopyrightpermission[1]{}    % 去掉版权声明
\setcopyright{none}    % 去掉版权信息

\acmDOI{none}
\acmISBN{none}
\acmConference{none}{none}{none}

\begin{document}
\title{Enhancing LLM-based Recommendation through Semantic-Aligned Collaborative Knowledge}

\author{Zihan Wang}
\authornotemark[1]
\affiliation{%
  \institution{Northeastern University}
  \city{Shen Yang}
  \country{China}
}
\email{2310744@stu.neu.edu.cn}

\author{Jinghao Lin}
\affiliation{%
  \institution{Northeastern University}
  \city{Shen Yang}
  \country{China}
}
\email{jinghao.lin@aminer.cn}

\author{Xiaocui Yang}
\affiliation{%
  \institution{Northeastern University}
  \city{Shen Yang}
  \country{China}
}
\email{yangxiaocui@cse.neu.edu.cn}

\author{Yongkang Liu}
\affiliation{%
  \institution{Northeastern University}
  \city{Shen Yang}
  \country{China}
}
\email{misonsky@163.com}

\author{Shi Feng}
\authornotemark[2]
\affiliation{%
  \institution{Northeastern University}
  \city{Shen Yang}
  \country{China}
}
\email{fengshi@cse.neu.edu.cn}

\author{Daling Wang}
\affiliation{%
  \institution{Northeastern University}
  \city{Shen Yang}
  \country{China}
}
\email{wangdaling@cse.neu.edu.cn}

\author{Yifei Zhang}
\affiliation{%
  \institution{Northeastern University}
  \city{Shen Yang}
  \country{China}
}
\email{zhangyifei@cse.neu.edu.cn}

\renewcommand{\shortauthors}{Trovato et al.}

\begin{abstract}
Large Language Models (LLMs) demonstrate remarkable capabilities in leveraging comprehensive world knowledge and sophisticated reasoning mechanisms for recommendation tasks. 
However, a notable limitation lies in their inability to effectively model sparse identifiers (e.g., user and item IDs), unlike conventional collaborative filtering models (Collabs.), thus hindering LLM to learn distinctive user-item representations and creating a performance bottleneck. 
Prior studies indicate that integrating collaborative knowledge from Collabs. into LLMs can mitigate the above limitations and enhance their recommendation performance. 
Nevertheless, the significant discrepancy in knowledge distribution and semantic space between LLMs and Collab. presents substantial challenges for effective knowledge transfer. 
To tackle these challenges, we propose a novel framework, SeLLa-Rec, which focuses on achieving alignment between the semantic spaces of Collabs. and LLMs. 
This alignment fosters effective knowledge fusion, mitigating the influence of discriminative noise and facilitating the deep integration of knowledge from diverse models. 
Specifically, three special tokens with collaborative knowledge are embedded into the LLM's semantic space through a hybrid projection layer and integrated into task-specific prompts to guide the recommendation process. 
Experiments conducted on two public benchmark datasets (MovieLens-1M and Amazon Book) demonstrate that SeLLa-Rec achieves state-of-the-art performance. 
\end{abstract}





%%
%% This command processes the author, affiliation and title
\maketitle

\section{Introduction}
In today’s era of information explosion, recommendation systems exhibit advanced capabilities in modeling user preferences and efficiently retrieving personally relevant content from vast information repositories \cite{bobadilla2013recommender,ko2022survey}. 
By leveraging data to understand both users and content, they play a pivotal role in establishing personalized connection mechanisms and have become a core component of modern intelligent systems. 
Traditional recommendation models mainly rely on collaborative patterns from user interactions—including clicks, ratings \cite{wu2022survey,koren2021advances}, and purchase behaviors—to build preference models (Collab.). 
Despite their success, these methods face inherent limitations tied to the availability and quality of user interaction data, especially in cold-start scenarios \cite{gope2017survey,lee2019melu} with minimal or no user history. 
The emergence of Large Language Models (LLMs) marks a transformative shift in recommendation systems research, offering novel approaches to overcome conventional recommendation frameworks' traditional constraints and performance limitations \cite{wu2024survey,zhao2024recommender}. 
LLMs offer three key strengths: their built-in knowledge for recommendation reasoning, enhanced performance through domain-specific fine-tuning, and effective handling of cold-start scenarios \cite{wang2024large,lin2024data}. 

However, being a general-purpose model, pure LLMs struggle to model extensive user and item IDs like recommendation-specialized collaborative recommendation systems (Collabs.), creating a performance bottleneck in recommendations \cite{zhang2025collm}. 
Recent research has begun exploring ways to integrate collaborative knowledge into LLMs as supplementary information, allowing LLMs to benefit from both LLM knowledge and collaborative knowledge sources. 
CoLLM \cite{zhang2025collm} pioneers this approach by projecting collaborative knowledge from pre-trained traditional models into LLMs' semantic space as specialized tokens within task prompts. Building on this foundation, BinLLM \cite{zhang2024text} further advances the field by encoding collaborative information as IP addresses and training models to interpret these addresses as user preference representations, achieving better performance. 
Generally speaking, these methods conceptualize LLMs and Collabs. as two independent systems, each grounded in distinct knowledge distributions. 
They aim to adopt collaborative signals into representations that are more interpretable within the semantic space of LLMs by mapping the knowledge from independently trained Collab..
Nevertheless, the significant disparity in knowledge distributions between the two types of models renders a single training of the projection layer insufficient for enabling LLMs to develop a comprehensive understanding of collaborative signals. 

To address this issue, we introduce SeLLa-Rec, which emphasizes the \textbf{Se}mantic Alignment between \textbf{LL}M and Coll\textbf{a}b. to enhance LLM understanding for collaborative knowledge. 
SeLLa-Rec consists of three layers: a \textbf{Collaborative Knowledge Foundation Layer} at the bottom, a \textbf{Hybrid Projection Layer} in the middle, and \textbf{LLM Recommendation Layer} at the top. 
Information flows upward through these layers, ultimately generating recommendations through the LLM layer. 

We initially enhance the top layer's LLM backbone through Lora \cite{hu2022lora} fine-tuning with recommendation-specific instructions, enabling it to better adapt to recommendation scenarios and achieve superior recommendation performance. 
Next, we distill items' semantic embeddings \cite{sheng2024language} using the fine-tuned LLM's knowledge. 
Within the collaborative knowledge foundation layer, these semantic embeddings serve as contrastive learning targets, enabling Collab. to pre-align with the LLM's semantic space during training. 
The pre-align approach allows the Collab. to inherit the LLM's knowledge distribution characteristics, establishing a solid foundation for subsequent knowledge projection. 
The alignment process operates bidirectionally, with the extracted semantic knowledge naturally assimilating aspects of collaborative knowledge during contrastive learning. This represents an active learning mechanism where LLM knowledge is enriched by collaborative insights. 
Consequently, we preserve the semantic knowledge obtained through the alignment process to complement Collab's collaborative knowledge. 
At the middle-tier Hybrid Mapping Layer, we integrate both the Collab. knowledge and the collaboratively-enhanced semantic knowledge rooted in LLM. 
This layer transforms these inputs into three specialized tokens that encapsulate collaborative information for LLM consumption. 
Notably, to strengthen the connection between LLM and Collab, we initialize the projection layer between them using parameters from the trained projection layer during the above collaborative model contrastive learning, rather than employing random initialization.  
It serves as a warm-up approach for the projection layer, minimizing information loss during dimensional transformation from lower to higher dimensions and enhancing overall mapping efficiency. 
Finally, with the LLM backbone frozen, we leverage carefully curated data incorporating three special tokens to train both the projection layers and downstream collaborative embeddings. 
This process can more effectively integrate collaborative knowledge into the LLM's semantic space, ultimately enhancing its comprehension and recommendation capabilities. 
SeLLa-Rec generally employs a hierarchical training approach similar to the sequence described above. 
It enables SeLLa-Rec to progressively transition from an initial, familiar recommendation task to more advanced recommendation generation by leveraging three specialized tags enriched with semantically consistent collaborative knowledge.
We summarize our contributions as follows: 
\begin{itemize}
    \item We present SeLLa-Rec, a recommendation model combining LLMs with collaborative knowledge through effective alignment of knowledge to improve recommendation quality. 
    \item We implement a hierarchical training strategy to guide the LLM to leverage special tokens with semantic-aligned collaborative knowledge, enhancing its recommendations. 
    \item We evaluate SeLLa-Rec extensively on MovieLens-1M and Amazon Book datasets, confirming its superior performance and the effectiveness of our alignment strategy. 
\end{itemize}

\section{Related Work}
The integration of Large Language Models (LLMs) into recommendation systems has garnered significant attention, primarily in two key directions: utilizing the extensive world knowledge of LLMs to help solve recommendation tasks and employing LLMs directly as standalone recommenders. 
\subsection{LLM-enhanced Recommendation}
Researchers first highlight whether LLMs could help improve the effectiveness of recommendations.
For example, studies such as SAID \cite{hu2024enhancing}, Recformer \cite{li2023text}, and HLLM \cite{chen2024hllm} have shown that leveraging LLM-derived embeddings enhances both the accuracy and relevance of recommendations. 
Additionally, LLMs have been employed as rerankers \cite{hou2024large,Sun2023InstructionDM} to further improve the accuracy of recommendation recall.  
Beyond these performance improvements, LLMs have also been applied to address various other challenges in recommendation systems. 
The remarkable role-playing capabilities of LLMs have also gained attention. 
With access to reliable user information, LLMs can simulate user behavior to generate synthetic behavior data, addressing issues such as data sparsity. 
This approach has been explored in studies like Agent4Rec \cite{zhang2024generative}, RecAgent \cite{wang2023user}, and Recommender AI Agent \cite{huang2023recommender}. 
The interpretability of recommendations has long been a key area of focus, as it helps to uncover the underlying logic of recommendation algorithms. 
LLMs hold significant potential for exploring the reasoning behind recommendation outcomes \cite{lei2024recexplainer,gao2023chat,ma2024xrec}. 

\subsection{LLM as Recommender}
LLMs, with their reasoning capabilities and strong generalization in recommendation tasks, are being explored as direct recommenders for the next-generation recommendation paradigm \cite{zhang2021language}. 
Broadly, using LLMs as recommenders can be categorized into two approaches. 
The first leverages the generative capabilities of LLMs to build generative recommendation models \cite{li2023large,deldjoo2024recommendation,wang2023generative}. 
The second reformulates recommendation tasks into textual formats, allowing LLMs to generate structured outputs to complete the task \cite{he2023large,lyu2023llm,kim2024large,liu2024cora,bao2023bi}. 
We primarily focus on the latter. 
Geng \cite{geng2022recommendation} proposes a unified recommendation framework, P5, that uses LLMs for pre-training. 
TallRec \cite{bao2023tallrec} introduces an instruction tuning method to boost LLM's recommendation performance. 
CoLLM \cite{zhang2025collm} and BinLLM \cite{zhang2024text}, which further explore the integration of collaborative filtering with LLMs. 
However, the unresolved semantic and knowledge distribution gap between these models impairs LLMs' understanding of collaborative knowledge, thereby limiting their performance. 
SeLLa-Rec is purposefully designed to bridge this gap. 

\section{Preliminaries}
\label{sec:preliminary}
Before detailing our proposed methodology, we present a comprehensive overview of the recommendation task and establish fundamental concepts for understanding the subsequent technical discussion. 

\textbf{Problem Formulation.} 
Building on the setups of CoLLM \cite{zhang2025collm} and BinLLM \cite{zhang2024text}, we adopt the click-through rate (CTR) prediction task \cite{yang2022click,zhang2021deep} as the focus of our study. 
Let $\mathcal{I}=\{i_1, i_2, ...,i_n\}$ denote the item set and $\mathcal{U}=\{u_1,u_2, ..., u_m\}$ denote the user set. 
For a user $u \in \mathcal{U}$ with an interaction history $\mathcal{H}_u$, the CTR task aims to predict the preference score $\hat{y}_{u,i} \in [0,1]$ for a candidate item $i \in \mathcal{I}$ based on $\mathcal{H}_u$, where the true label $y_{u,i} = 0~or ~1$. 
A value of $\hat{y}{u,i}$ closer to 1 indicates a higher likelihood that user $u$ would enjoy item $i$, and the closer $\hat{y}_{u,i}$ is to $y_{u,i}$, the more accurate the prediction. 

\textbf{Collaborative Filtering Recommendation.} 
Collaborative filtering models (Collab.) \cite{zhou2018deep,xiao2020deep} explicitly construct user preferences and item features as learnable embeddings 
$ \mathbf{E}^C_U = \{\mathbf{e}^{C}_{u_1}, \mathbf{e}^{C}_{u_2}, ...\mathbf{e}^{C}_{u_m}\}$ and $\mathbf{E}^C_I = \{\mathbf{e}^{C}_{i_1}, \mathbf{e}^{C}_{i_2}, ...\mathbf{e}^{C}_{i_n}\}$, where $\mathbf{e}^{C}_{u}, \mathbf{e}^{C}_{i} \in \mathbb{R}^{d_C}$ and $d_C$ typically ranges from 64 to 256. 
The predicted preference score $\hat{y}_{u,i}$ can be computed as $f^C(\mathbf{e}^{L}_{u}, \mathbf{e}^{L}_{i})$  
where $f^C$ is either inner product $<\mathbf{e}^{C}_{u}, \mathbf{e}^{C}_{i}>$ or other neural network. 

\begin{figure*}[htbp]
    \centering
    \includegraphics[width=0.97\textwidth]{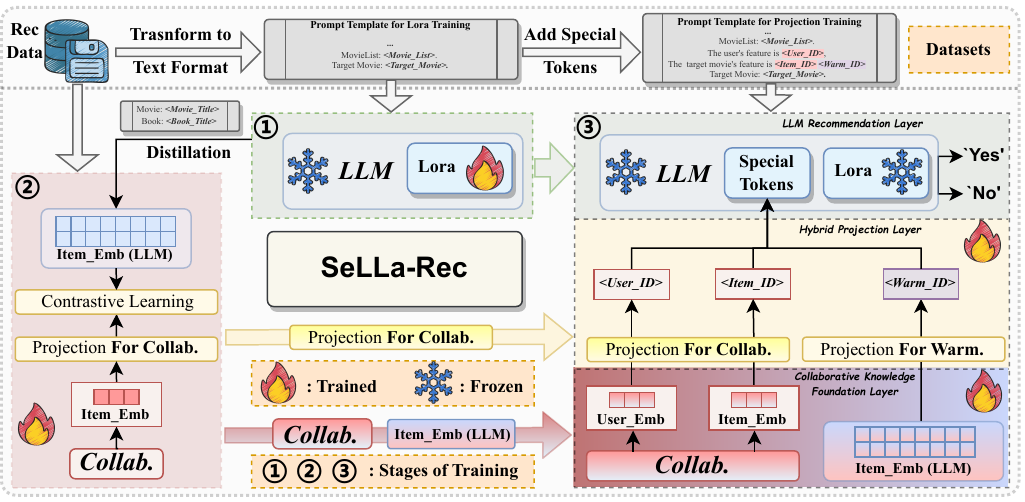}
    \caption{The overall framework of SeLLa-Rec. 
    The right part depicts SeLLa-Rec's structure, while the left illustrates the origins of the pre-trained components. 
    The numbered markers \ding{172}, \ding{173}, and \ding{174} denote the sequential training stages for SeLLa-Rec. }
    \label{fig:framework}
\end{figure*}
\textbf{LLM-based Recommendation.} 
To enable the LLM to perform the CTR task, we describe the task along with the user interaction history $\mathcal{H}_u$ and the target item $i$, both converted into text format and supplemented with text prompts, as the model's input $\mathcal{P}^{L}$. 
The LLM first tokenizes $\mathcal{P}^{L}$ and then embeds these tokens into its own representation space, obtaining $\mathbf{E}^L(\mathcal{P})$. 
During embedding, each token is transformed into a tensor $e^L$. Specifically, $e^L \in \mathbb{R}^{d_L}$, where the dimension $d_L$ typically falls within the range of 3072 to 4096 and is substantially larger than $d_C$. 

Subsequent LLM layers \( f^L \) process these embeddings to generate token-level logits \( \mathbf{l} \in \mathbb{R}^{|\mathcal{V}|} \) across the model’s vocabulary \( \mathcal{V} \). 
The final logits calculated by LLM can be simply expressed as follows: 
\begin{equation}
    \mathbf{Logits} = f^L(E^L(\mathcal{P}^{L}))
\end{equation}
Then the logits corresponding to the `\texttt{Yes}' and `\texttt{No}' tokens at the last prediction position is extracted in Equ.~\ref{equ:yes_no}:
\begin{equation}
\label{equ:yes_no}
    [l_{\text{Yes}}, l_{\text{No}}] = \mathbf{Logits} \cdot [e(v_{\text{Yes}}), e(v_{\text{No}})]^\top
\end{equation}
$e(v_{\text{Yes}})$ and $e(v_{\text{No}})$ denote the one-hot encoded vectors corresponding to the vocabulary tokens `\texttt{Yes}' and `\texttt{No}' respectively, where the vectors are mapped from their corresponding token IDs in vocabulary $\mathcal{V}$. 
The preference probability \( \hat{y}_{u,i} \) is derived via a constrained softmax operation: 
\begin{equation}
\label{equ:y_hat}
    \hat{y}_{u,i} = \exp(l_{\text{Yes}}) / \left(\exp(l_{\text{Yes}}) + \exp(l_{\text{No}}) \right) 
\end{equation}
This probability directly quantifies the user’s likelihood of interacting with the target item, repurposing the LLM’s generative space into a discriminative scoring for recommendation \cite{bao2023tallrec}. 

\section{Methodology}
\label{sec:method}
SeLLa-Rec composes three layers, as illustrated in Fig~\ref{fig:framework}. 
The top layer is the \textbf{LLM recommendation layer}, serving as the backbone of the entire model to generate final recommendations. 
The bottom layer is the \textbf{Collaborative Knowledge Foundation Layer}, which provides collaborative signals for LLM. 
The middle layer is the \textbf{Hybrid Projection Layer}, which receives text input and collaborative information from the bottom layer and encodes them into tokens for LLM understanding. 
In this section, we provide detailed explanations of these three layers' composition and training methods with mathematical formulas. 

\begin{figure}[htbp]
    \centering
    \includegraphics[width=0.4\textwidth]{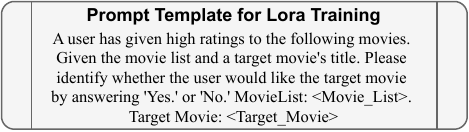}
    \caption{The prompt structure for Lora finetuning.}
    \label{fig:prompt_1}
\end{figure}
\subsection{LLM Recommendation Layer}
\label{sec:llm_rec_layer}
The LLM recommendation layer is the model's core architecture and ultimate decision-making component. 
Its functionality can be dissected into two critical aspects: comprehending the recommendation task and interpreting collaborative knowledge. 
To address the first aspect, we adopt TaLLRec's \cite{bao2023tallrec} approach of encoding recommendation tasks in textual format and fine-tuning the LLM, which we elaborate on in this subsection. 
The second aspect is accomplished through training in the heterogeneous mapping layer, which we detail in Section.~\ref{sec:projection}. 

To align the LLM's generation capability with recommendation objectives, we leverage lightweight Low-Rank Adaptation (Lora) \cite{hu2022lora} to perform supervised fine-tuning (SFT) following the prompt $\mathcal{P}^L$. 
$\mathcal{P}^L$ is constructed as shown in Fig.~\ref{fig:prompt_1}, incorporating task descriptions and relevant user and item information, which is transformed into text. 
Following the procedure outlined in Section.~\ref{sec:preliminary}, we compute the prediction $\hat{y}_{u,i}$ of LLM with Lora parameters $\Theta_{\text{Lora}}$ for training as Equ.~\ref{equ:lora}. 
\begin{equation}
\label{equ:lora}
    \mathbf{Logits} = f^L(E^L(\mathcal{P});\Theta_{\text{Lora}})
\end{equation}

The golden response (label) is a binary word (`\texttt{Yes}'/`\texttt{No}') according to $y_{u,i}$
The training objective is to minimize the expectation $\mathbb{E}$ of the binary cross-entropy loss as Equ.~\ref{equ:loss_sft}:
\begin{equation}
\label{equ:loss_sft}
    \mathcal{L}_{\text{SFT}} = -\mathbb{E} \left( y_{u,i} \log \hat{y}_{u,i} + (1-y_{u,i}) \log(1-\hat{y}_{u,i}) \right )
\end{equation}

\begin{figure}[htbp]
    \centering
    \includegraphics[width=0.4\textwidth]{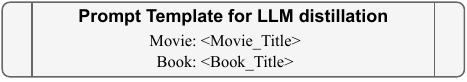}
    \caption{The prompt structure for LLM distillation.}
    \label{fig:prompt_2}
\end{figure}
\subsection{Collaborative Knowledge Foundation Layer}
\label{sec:collab}
This layer of collaborative knowledge is built upon two fundamental components: Collab. and the semantic embeddings distilled from the LLM. 
During the training process, these semantic embeddings guide the Collab. to align with the LLM's semantic space. 
Through the bidirectional alignment, the semantic embeddings become enriched with collaborative knowledge. 
Ultimately, after training, both components function as complementary sources of collaborative information, feeding the collaborative knowledge to the LLM. 
The details are illustrated below. 

As the first step, we distill the knowledge of the LLM equipped with the trained Lora module in Section.~\ref{sec:llm_rec_layer}. 
Specifically, we construct a prompt for each item as Fig.~\ref{fig:prompt_2} and take the embedding of the next token as the item's semantic embedding from LLM like \cite{sheng2024language} to get $E^L_I = \left\{e^L_{i_1}, e^L_{i_1},...,e^L_{i_n} \right\}$. 
Then, we train the downstream Collab.. 
Traditional collaborative filtering (Collab.) models learn user and item embeddings ($E^C_U$ and $E^C_I$) from user-item interaction matrices. 
These models cluster users with similar behaviors and items with related features in the embedding space. 
The prediction process can be formulated through these learned embeddings, with the training objective defined as: 
\begin{equation}
\mathcal{L}_{\text{collab}} = - \mathbb{E} \left( y_{u,i} - (\mathbf{e}^{C}_{u})^{\top} \mathbf{e}^{C}_{i} \right)^2
\end{equation}
where $\mathbf{e}^{C}_{u} \in \mathbb{R}^{d_C}$ and $\mathbf{e}^{C}_{i} \in \mathbb{R}^{d_C}$ denote user and item embeddings from collaborative filtering models respectively. 

To achieve proper alignment between embeddings learned during Collab. training and those semantic embeddings from LLM, we introduce an additional alignment loss term $\mathcal{L}_{\text{align}}$ using contrastive learning (InfoNCE) \cite{oord2018representation} as demonstrated by Equ.~\ref{equ:cl}. 
Given the substantial dimensional disparity between the collaborative model and LLM embeddings, we employ a dimension mapping function $Proj^{C \rightarrow L}$ in Equ.~\ref{equ:proj_collab} to project the collaborative embeddings into the LLM's embedding space. 
\begin{equation}
\label{equ:proj_collab}
Proj^{C \rightarrow L}(\mathbf{e}_i^{\text{C}}) = \mathbf{W}_2(\text{GELU}(\mathbf{W}_1\mathbf{e}_i^{\text{C}} + \mathbf{b}_1)) + \mathbf{b}_2
\end{equation} 
\begin{equation}
\label{equ:cl}
\mathcal{L}_{\text{align}} = - \mathbb{E} \left(  \frac{\exp\left(\cos(Proj^{C \rightarrow L}(\mathbf{e}^{C}_{i}), \mathbf{e}^{L}_{i})/\tau\right)}{\sum_{i' \in \mathcal{I}} \exp\left(\cos(Proj^{C \rightarrow L}(\mathbf{e}^{C}_{i'}), \mathbf{e}^{L}_{i'})/\tau\right)} \right)
\end{equation}
$\tau$ is the temperature for InfoNCE. 
Therefore, the Collab. for SeLLa-Rec maintains a persistent hybrid training objective: 
\begin{equation}
\label{equ:stage2}
\mathcal{L} = \mathcal{L}_{\text{collab}} + \lambda \mathcal{L}_{\text{align}}
\end{equation}

Finally, we get three useful components for future training: 
(1) Trained embeddings from Collab. ${E^{C}}' = \left\{{E^{C}_{U}}', {E^{C}_{I}}'\right\}$. 
(2) Items' semantic embeddings after alignment ${E^L_I}'$. 
(3) The pretrained projection network ${Proj^{C \rightarrow L}}'$ bridging the above two embeddings. 
\begin{figure}[htbp]
    \centering
    \includegraphics[width=0.4\textwidth]{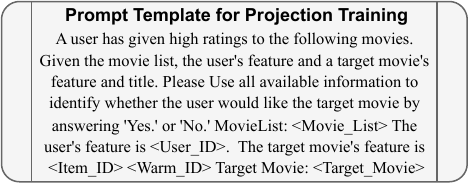}
    \caption{The prompt structure for Projection Training. }
    \label{fig:prompt_3}
\end{figure}

\subsection{Hybrid Projection Layer}
\label{sec:projection}
The Hybrid Projection Layer serves as a crucial bridge between the core components of SeLLa-Rec, projecting collaborative information into the LLM's semantic space. 
This layer incorporates two specialized projection modules that correspond to the two key components from the Collaborative Foundation Layer: 
(1) Collab.'s embeddings for users and items ${E^{C}}' = \left\{{E^{C}_{U}}', {E^{C}_{I}}'\right\}$, and 
(2) Items' semantic embeddings after bidirectional alignment ${E^{L}}' = \left\{{E^{L}_{I}}'\right\}$. 
These three types of embeddings $\left\{{E^{C}_{U}}', {E^{C}_{I}}', {E^{L}_{I}}'\right\}$ directly align with the special tokens provided to the LLM, $<User\_ID>$, $<Item\_ID>$, and $<Warm\_ID>$. 
They are designated placeholders within the LLM prompt, constituting $\mathcal{P}^{L + C}$ as Fig.~\ref{fig:prompt_3}. 
During the LLM encoding process, the embeddings at these three token positions undergo substitution with embeddings derived from mapped collaborative knowledge, thereby facilitating the seamless integration of collaborative information into LLM. 
The first two embeddings, with the dimension $d_C$, follow the Collab. distribution, so share a common projection layer $Proj^{C \rightarrow L}$ and third embedding, with dimension $d_L$, utilizes a separate projection layer $Proj^{W\rightarrow L}$ as Equ.~\ref{equ:tokens}. 
\begin{itemize}
\label{equ:tokens}
    \item $<User\_ID>$: $e^L_{user} = Proj^{C \rightarrow L}(\mathbf{e}^{C}_{u})$ captures user collaborative features. 
    \item $<Item\_ID>$: $e^L_{item} =Proj^{C \rightarrow L}(\mathbf{e}^{C}_{i})$ captures item collaborative features.
    \item $<Warm\_ID>$: $e^L_{warm} = Proj^{W \rightarrow L}( \mathbf{e}_i^{L})$ preserves aligned LLM knowledge for item features. 
\end{itemize} 
The projection network parameters denoted by $Proj^{C \rightarrow L}$ are established during the Collab. training in Section~\ref{sec:collab}. 
We adopt this approach based on the understanding that this layer has already accomplished the necessary semantic alignment, thus facilitating more effective semantic space alignment with the LLM. 
In contrast, $Proj^{W \rightarrow L} \in \mathbb{R}^{d_L \times d_L}$ represents a learnable linear projection layer that begins with random initialization. 
These transformed embeddings of $\mathcal{P}^{L + C}$ are in Equ.~\ref{equ:transformed_emb}. 
\begin{equation}
\label{equ:transformed_emb}
    E^L_{final} = E^L(\mathcal{P}^{L+C}) \oplus [e^L_{user};e^L_{item};e^L_{warm}]
\end{equation}
The training objective combines recommendation loss with representation fidelity constraints:
\begin{equation}
    \mathcal{L}_{\text{Proj}} = - y \log \hat{y}^{L+C}_{u.i} + (1-y) \log(1-\hat{y}^{L+C}_{u.i})
\end{equation}
where $\hat{y}^{L+C}_{u.i} = p(\hat{y}=1|\left\{f^L,E^L_{final}\right\})$. 

\begin{table}[t]
\centering
\caption{Stage-wise Component Training Status}
\label{tab:training_status}
\begin{small}
\begin{tabular}{l|ccc}
\hline
\textbf{Component} & Stage 1 & Stage 2 & Stage 3 \\ \hline
LLM Backbone & Frozen & Frozen & Frozen \\
LoRA Parameters $\Theta_{\text{Lora}}$ & Trainable & Frozen & Frozen \\
Collab. Embeddings $E^C = \left\{E^C_U, E^C_I\right\}$ & - & Trainable & Trainable \\
LLM Item Embeddings $E^L= \left\{E^L_I\right\}$ & - & Trainable & Trainable \\
Collab. Projection Layer $Proj^{C \rightarrow L}$ & - & Trainable & Trainable \\
Warm. Projection Layer $Proj^{W \rightarrow L}$ & - & - & Trainable \\ \hline
\end{tabular}
\end{small}
\end{table}

\subsection{Training Procedure}
Following our discussion of component-specific training methods, this subsection provides a holistic overview of our complete training process. 
Our training process consists of three sequential stages illustrated in  Tab.~\ref{tab:training_status}: 

\textbf{Stage 1}: Task Adaptation via LoRA fine-tuning to align the LLM Recommendation Layer backbone with recommendation objectives through supervised learning on formatted prompts.

\textbf{Stage 2}: The Collaborative Knowledge Foundation layer undergoes joint training for dual objectives: recommendation and alignment tasks with distilled knowledge from the LLM. 

\textbf{Stage 3}: The training process simultaneously optimizes both the Hybrid Projection layer's two projection components and the Collaborative Knowledge established in stage 2, facilitating improved knowledge integration with the LLM. All LLM-related parameters, including those in the LoRA modules, are frozen during this stage. 

\section{Experiment}
In this section, we present experimental results and related analysis. 
We begin by outlining the basic experimental settings. 
Next, we compare the performance of SeLLa-Rec with several traditional and/or state-of-the-art models and analyze how SeLLa-Rec performs across different types of data. 
Lastly, we conduct an in-depth investigation into the contribution of SeLLa-Rec's components.
\subsection{Experimental Settings}
\textbf{Datasets.} To ensure a rigorous and equitable comparative evaluation, our experiments are all conducted on two public datasets to maintain consistency with CoLLM \cite{zhang2025collm} and BinLLM \cite{zhang2024text}. 
\textbf{MovieLens-1M} \cite{harper2015movielens} (Movie for short) is from the widely-recognized Movie\footnote{\url{https://grouplens.org/datasets/movielens/1m/}} dataset, a comprehensive collection of user-movie interactions gathered between 2000 and 2003. 
It comprises user ratings on a five-point scale (1-5), which are subsequently transformed into binary classifications using a threshold value of 3. 
Specifically, ratings exceeding 3 are designated positive interactions, while ratings of 3 and below are categorized as negative interactions. 
\textbf{Amazon-Book} \cite{hou2024bridging} (Book for short) refers to the "Book" subset of the well-known Amazon Product Review dataset\footnote{\url{https://cseweb.ucsd.edu/~jmcauley/datasets.html\#amazon_reviews}} . 
It compiles user reviews of books on Amazon, collected between 1996 and 2018, with review scores ranging from 1 to 5. 
These review scores are converted into binary labels using a threshold of 4. 

Datasets are partitioned into training, validation, and test sets based on temporal order to better reflect real-world scenarios.
Movie dataset stores interactions from the last 20 months, and is split chronologically in a 10:5:5 ratio for training, validation, and testing.
Due to its large size, the Book dataset uses only interactions from 2017. To ensure data quality and reliability, only users with more than 20 interactions are included. 
The data is then divided chronologically in an 11:0.5:0.5 ratio. 
The time-based division method ensures that the test set includes items not appearing in earlier periods, effectively simulating a natural cold-start scenario.
Dataset statistics are shown in Tab.~\ref{tab:dataset_statistics}. 

\begin{table}[htbp]
\caption{Statistics of datasets.}
\label{tab:dataset_statistics}
\centering
\resizebox{0.45\textwidth}{!}{
\begin{threeparttable}
\begin{small}
\begin{tabular}{c|cccccc}
\toprule
\textbf{Dataset} & \textbf{\#Train} & \textbf{\#Valid} & \textbf{\#Test} & \textbf{\#User} & \textbf{\#Item} \\ 
\midrule
MovieLens-1M           & 33,891           & 10,401           & 7,331           & 839             & 3,256           \\ 
Amazon-Book     & 727,468          & 25,747           & 25,747          & 22,967          & 34,154          \\ 
\bottomrule
\end{tabular}
\end{small}
\end{threeparttable}
}
\end{table}

\textbf{Baselines.} We select a total of eleven baseline models spanning four distinct categories for a comprehensive comparison. 
Below is a brief introduction to the baseline models:

\noindent Traditional collaborative models (Collab. for short):
\begin{itemize}
    \item \textbf{MF} \cite{koren2009matrix} is a classic recommendation method based on the matrix decomposition.
    \item \textbf{LightGCN} \cite{he2020lightgcn} represents the typical graph neural network (GNN)-based collaborative filtering method.
    \item \textbf{SASRec} \cite{kang2018self} leverages self-attention mechanisms to make sequential recommendation model.
    \item \textbf{DIN} \cite{zhou2018deep} is a classic click-through rate (CTR) prediction model that incorporates dynamic interest modeling. 
\end{itemize}

\noindent LLM-based recommendation models without collaborative knowledge (LLMREC for short): 
\begin{itemize}
    \item \textbf{Zero-shot LLM} directly utilizes the native capabilities of LLMs for recommendation without any fine-tuning. 
    \item \textbf{Prompt4NR} \cite{zhang2023prompt} employs prompt-based learning techniques to guide the model in generating recommendation outputs. 
    \item \textbf{TALLREC} \cite{bao2023tallrec} enhances the recommendation performance of LLMs through carefully designed instruction tuning.
\end{itemize}

\noindent Collabs. that leverage the knowledge of LLMs as auxiliary support (Collab. with LLM for short): 
\begin{itemize}
    \item \textbf{CTRL} \cite{li2023ctrl} integrates LLMs to embed tabular data for aiding the training of downstream CTR models. 
\end{itemize}

\noindent LLM-based recommendation models enhenced with collaborative information (LLMREC with Collab. for short):
\begin{itemize}
    \item \textbf{PersonPrompt} \cite{li2023personalized} introduces new tokens and token embeddings to incorporate collaborative signals, functioning as a personalized soft prompt.
    \item \textbf{CoLLM} \cite{zhang2025collm} maps the collaborative knowledge from the downstream Collab. directly into the LLM semantic space and integrates it into the prompt as additional tokens, enabling the LLM to comprehend the collaborative information. 
    \item \textbf{BinLLM} \cite{zhang2024text}, building on CoLLM, converts the knowledge from the downstream Collab. into a textual format (IP addresses), allowing the LLM to interpret collaborative signals in a text-based representation. 
\end{itemize}

\textbf{Metrics and other settings. } We use two commonly used metrics: AUC \cite{lobo2008auc} (Area Under the ROC Curve) and UAUC \cite{bao2023tallrec} (User Average AUC). 
AUC is primarily used to evaluate the overall prediction accuracy, while UAUC provides deeper insights into the recommendation quality at the user-level. 

Existing LLMREC models with Collab. architectures typically select the Vicuna-7B \cite{zheng2023judging} model as the backbone for the overall recommendation system, owing to its strong performance. 
Specifically, Vicuna-7B was previously considered a strong-performing model of its size. However, significant advancements in LLM performance have been achieved over the past two years.
According to the latest records from Huggingface's OpenLLM leaderboard\footnote{\url{https://huggingface.co/spaces/open-llm-leaderboard/open_llm_leaderboard\#/}}, Vicuna-7B has now fallen behind other models of the same size in terms of performance. 
Therefore, in our experiments, we replace all LLM backbones involved with the more advanced Qwen2-7B-Base \cite{qwen2}\footnote{\url{https://huggingface.co/Qwen/Qwen2-7B}} model.  
For downstream Collab.,  we select MF due to its simplicity. 
Unlike more complex models, MF does not include additional network structures to store collaborative knowledge, relying solely on the embeddings of users and projects. 
Moreover, practical experiences from CoLLM and BinLLM suggest that MF is stable as a downstream Collab., making it a suitable choice for the study. 

\begin{table}[htbp]
\caption{Overall performance and SeLLa-Rec and baselines while best results are in bold; second best are underlined.}
\label{tab:overall_performance}
\centering
\resizebox{0.48\textwidth}{!}{
\begin{threeparttable}
\begin{small}
\begin{tabular}{cc|ll|ll}
\toprule
\multicolumn{2}{c|}{\textbf{Datasets}}                                                                                                                                 & \multicolumn{2}{c|}{\textbf{Movie}}        & \multicolumn{2}{c}{\textbf{Book}}          \\ \midrule
\multicolumn{2}{c|}{\textbf{Models}}                                                                                                                                   & AUC             & UAUC            & AUC             & UAUC            \\ \midrule
\multicolumn{1}{c|}{\multirow{4}{*}{\textbf{Collab.}}}                                                          & MF                                                   & 0.6473          & 0.6264          & 0.6949          & 0.5366          \\
\multicolumn{1}{c|}{}                                                                                  & LightGCN                                             & 0.6139          & 0.6276          & 0.7011          & 0.5528          \\
\multicolumn{1}{c|}{}                                                                                  & SASREC                                               & 0.7058          & 0.6818          & 0.6721          & 0.5597          \\
\multicolumn{1}{c|}{}                                                                                  & DIN                                                  & 0.7138          & 0.6595          & 0.8102          & 0.5986          \\ \midrule
\multicolumn{1}{c|}{\multirow{3}{*}{\textbf{LLMREC}}}                                                           & Zero-shot LLM                                        & 0.5718          & 0.5145          & 0.6791          & 0.5546          \\
\multicolumn{1}{c|}{}                                                                                  & Prompt4NR                                            & 0.7191          & 0.6839          & 0.8177          & 0.6733          \\
\multicolumn{1}{c|}{}                                                                                  & TALLREC                                              & 0.7268          & 0.6989          & 0.8358          & 0.6921          \\ \midrule
\multicolumn{1}{c|}{\begin{tabular}[c]{@{}c@{}}\textbf{LLM-Collab.}\end{tabular}}                        & \begin{tabular}[c]{@{}c@{}}CTRL(DIN)\end{tabular} & 0.7155          & 0.6681          & 0.8122          & 0.6177          \\ \midrule
\multicolumn{1}{c|}{\multirow{3}{*}{\begin{tabular}[c]{@{}c@{}}\textbf{LLMREC}\\ \textbf{with} \\ \textbf{Collab.}\end{tabular}}} & PersonPrompt                                         & 0.7223          & 0.6991          & 0.8141          & 0.6730          \\
\multicolumn{1}{c|}{}                                                                                  & CoLLM-MF                                                & 0.7357          & 0.7179          & 0.8391          & 0.6957          \\
\multicolumn{1}{c|}{}                                                                                  & BinLLM                                               & {\ul 0.7417}    & {\ul 0.7239}    & {\ul 0.8429}    & {\ul 0.7073}    \\ \midrule
\multicolumn{1}{c|}{\textbf{ours}}                                                                              & SeLLa-Rec                                             & \textbf{0.7606} & \textbf{0.7464} & \textbf{0.8837} & \textbf{0.7459} \\ \bottomrule
\end{tabular}
\end{small}
\end{threeparttable}
}
\end{table}

\subsection{Overall performance}
Tab.~\ref{tab:overall_performance} presents the performance comparison between SeLLa-Rec and other baseline models on the Movie and Book datasets. 
The results in the table reveal four main results: 
(1) SeLLa-Rec outperforms all baseline models on both datasets. 
An average improvement of approximately 4\% is achieved on both datasets. 
Since AUC and UAUC values closer to 1 are more challenging to improve, this underscores the performance contribution of SeLLa-Rec. 
(2) It is evident that recommendation models based on LLMs consistently outperform traditional collaborative filtering models. 
It highlights the potential of utilizing LLMs for recommendation tasks. 
The performance of the LLMREC models reveals that zero-shot LLMs already exhibit strong recommendation capabilities, which can be significantly improved with some methods. This improvement is substantial. 
On one hand, this demonstrates that the inherent reasoning abilities of LLMs can be effectively applied to recommendation tasks. 
On the other hand, it suggests that these abilities may not yet be fully utilized, leaving ample room for further exploration. 
(3) CTRL leverages LLM knowledge to enhance the performance of the smaller model, DIN. It demonstrates that using LLM knowledge to guide the training of collaborative models is a reliable approach, aligning with our setting for training downstream Collab.. 
(4) Compared to basic LLMREC models, incorporating collaborative information (LLMREC with Collab.) generally improves performance. 
It represents a fusion of two types of knowledge, highlighting LLMs' sensitivity to external knowledge and their ability to incorporate and utilize it effectively.
However, PersonPrompt is an exception, as it does not leverage user and item embeddings from a well-trained collaborative model. 
Instead, it trains embeddings of special user and item tokens from scratch, which is less efficient. 

\begin{figure}[htbp]
    \centering
    \begin{minipage}{\linewidth}
        \centering
        \begin{minipage}{0.48\linewidth}
            \centering
            \includegraphics[width=\linewidth]{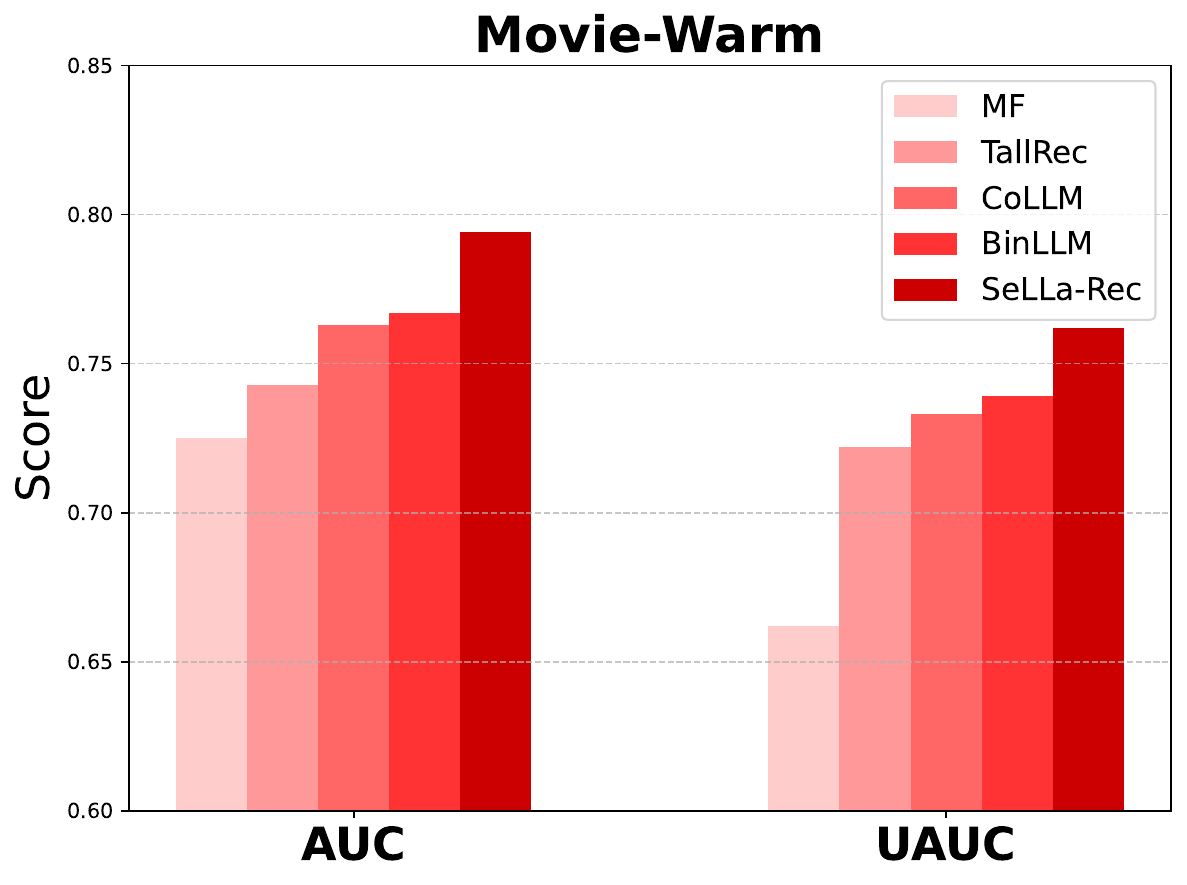} 
        \end{minipage}%
        \hfill
        \begin{minipage}{0.48\linewidth}
            \centering
            \includegraphics[width=\linewidth]{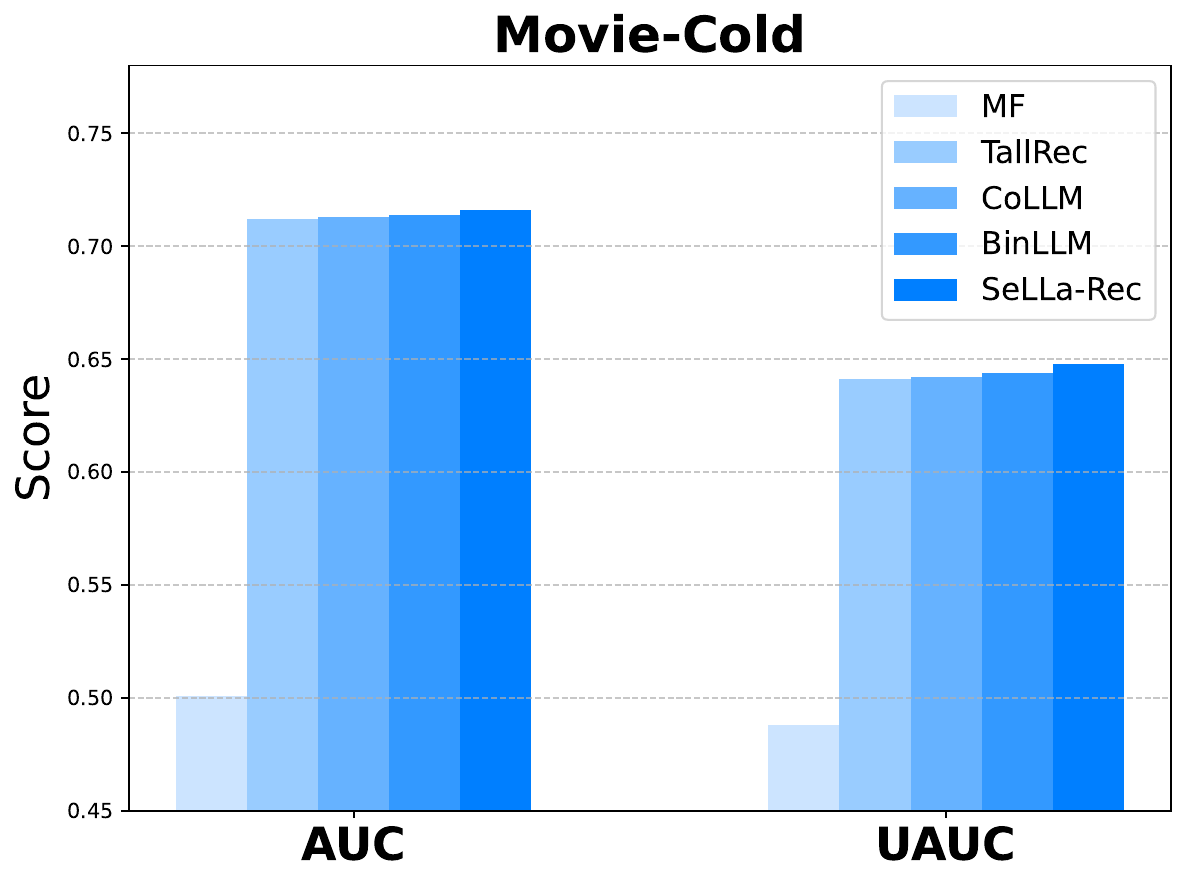} 
        \end{minipage}
    \end{minipage}%

    \vspace{1em} 

    \begin{minipage}{\linewidth}
        \centering
        \begin{minipage}{0.48\linewidth}
            \centering
            \includegraphics[width=\linewidth]{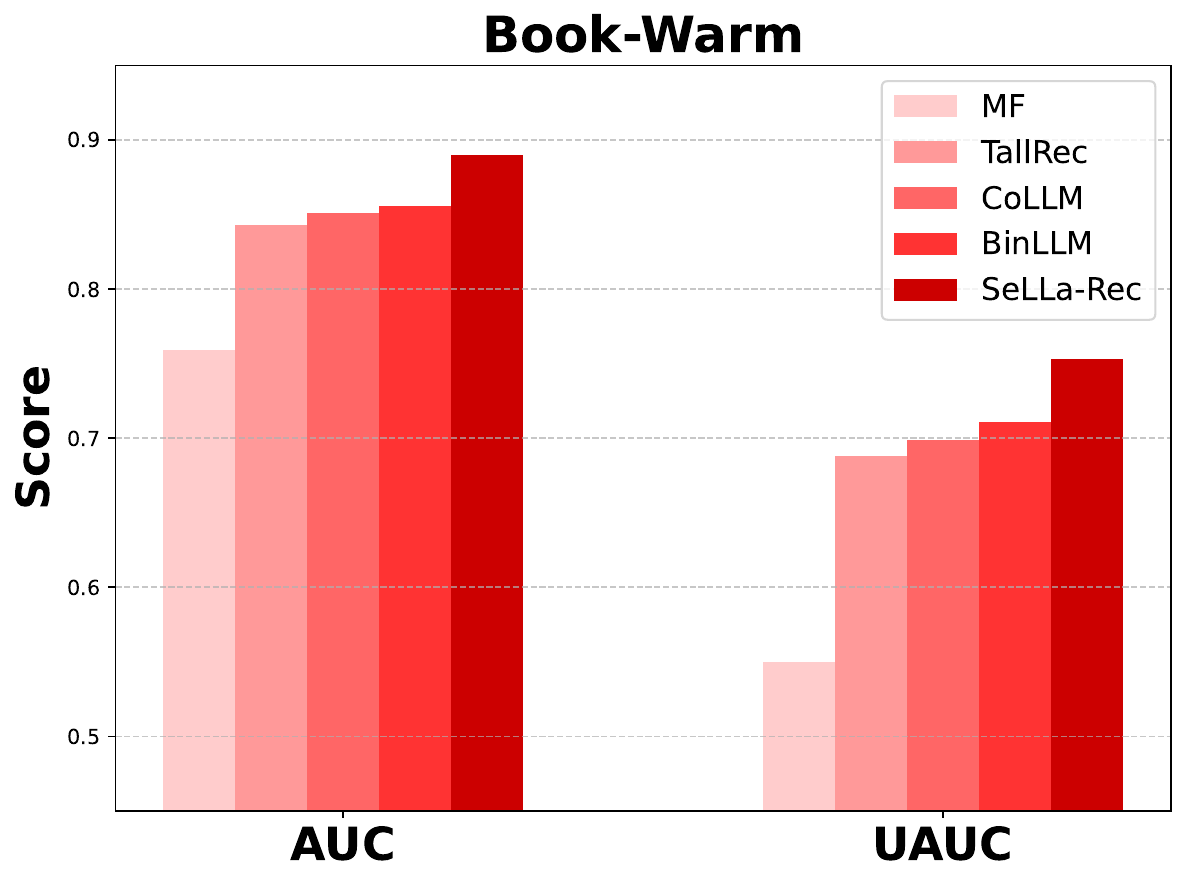} 
        \end{minipage}%
        \hfill
        \begin{minipage}{0.48\linewidth}
            \centering
            \includegraphics[width=\linewidth]{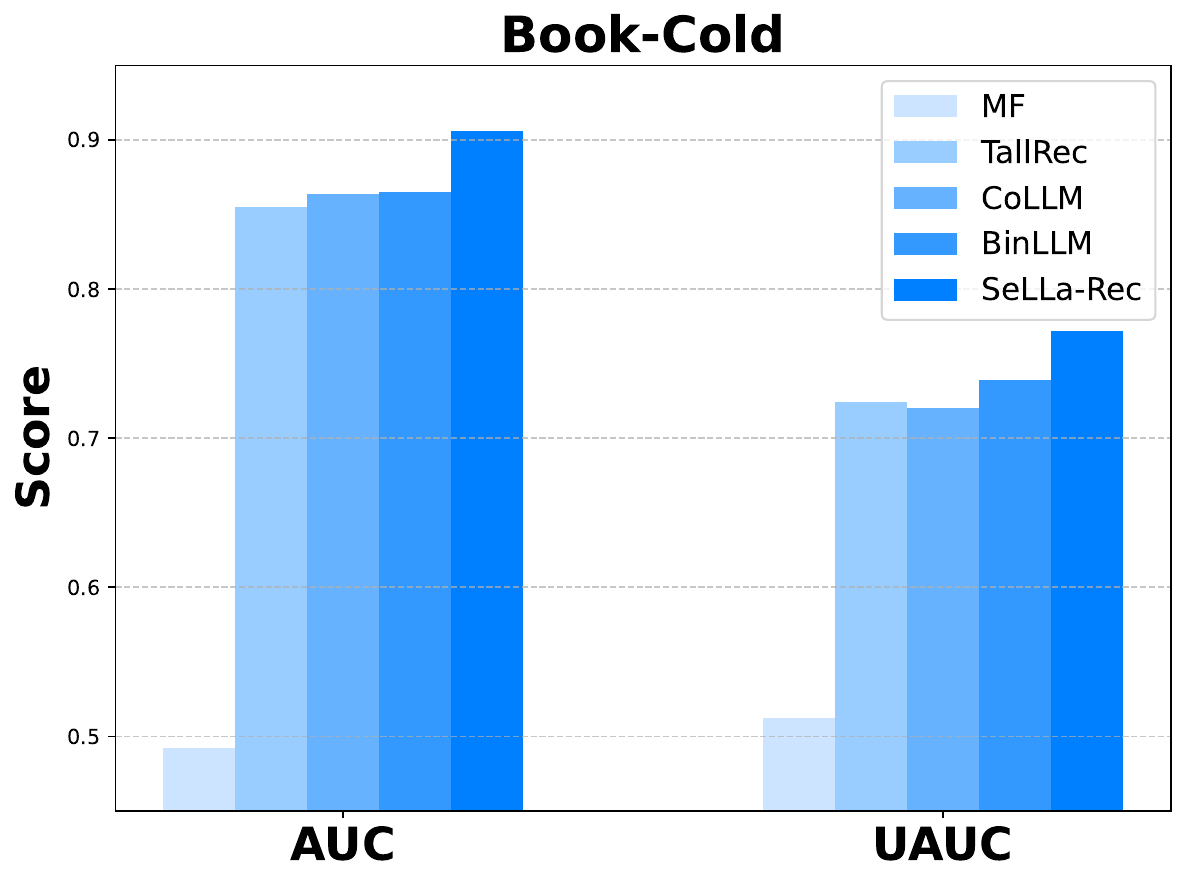} 
        \end{minipage}
    \end{minipage}%

    \caption{Models' performance on warm/cold data.}
    \label{fig:warm_cold}
\end{figure}

\subsection{Performance on Warm/Cold data}
Traditional collaborative models (Collabs.) are limited to recommending products and users seen in the training set (warm data), with performance dropping sharply on new, unseen data (cold data). 
It is particularly problematic given the rapid evolution of real-world products. 
Early studies on using LLMs for recommendation have therefore focused heavily on their ability to address the cold-start problem effectively. 
Research \cite{sanner2023large,wang2024large,zhang2025collm,lin2024data} has shown that LLM-based recommendation models, which rely solely on the reasoning capabilities of LLMs (LLMREC), outperform traditional collaborative models significantly on cold-start data. 
This strength comes from LLMs' ability to generalize using their prior knowledge of products. However, their performance on warm data (i.e., products and users present in the training set) often falls short of traditional collaborative models. 
Integrating collaborative information into LLM-based systems offers a promising approach to improving their performance on warm data \cite{zhang2025collm,zhang2024text}. 
To understand the source of SeLLa-Rec's improved recommendation performance, we evaluated it alongside baselines on both cold and warm data. 
As shown in Fig.~\ref{fig:warm_cold}, SeLLa-Rec achieves comprehensive improvements, enhancing performance on both data types. 

Specially, the performance of the LLM-based models on the Book dataset highlights two findings:
These models perform worse on the entire test set compared to both the hot and cold subsets, and they achieve better performance on the cold subset than on the hot subset.
This disparity may stem from the uneven data distribution and the presence of more familiar books in the test set \cite{davis2006relationship}.
To further investigate, we evaluate the LLM's recommendation ability in a zero-shot setting, and the results in Tab. \ref{tab:zero_shot_book_cold} confirm that the models perform much better on the cold data subset, supporting our hypothesis. 
It is worth emphasizing that this observation does not impact the experiment results. 

\begin{table}[htbp]
\caption{Zero-shot LLM's performance on Book's data.}
\label{tab:zero_shot_book_cold}
\centering
\resizebox{0.42\textwidth}{!}{
\begin{threeparttable}
\begin{small}
\begin{tabular}{c|ccc|ccc}
\toprule
\multicolumn{1}{c|}{} & \multicolumn{3}{c|}{\textbf{AUC}} & \multicolumn{3}{c}{\textbf{UAUC}} \\ 
\midrule
\textbf{Book Data} & \textbf{ALL} & \textbf{Warm} & \textbf{Cold} & \textbf{ALL} & \textbf{Warm} & \textbf{Cold} \\ 
\midrule
\textbf{Zero-shot LLM} & 0.6791 & 0.6602 & \textbf{0.7604} & 0.5546 & 0.5507 & \textbf{0.5643} \\ 
\bottomrule
\end{tabular}
\end{small}
\end{threeparttable}
}
\end{table}

\begin{table}[htbp]
\caption{Performance comparison for SeLLa-Rec and variants. }
\label{tab:ablation}
\centering
\resizebox{0.48\textwidth}{!}{
\begin{threeparttable}
\begin{small}
\begin{tabular}{cc|ll|ll}
\toprule
\multicolumn{2}{c|}{\textbf{Datasets}}                                                                                        & \multicolumn{2}{c|}{\textbf{Movie}}         & \multicolumn{2}{c}{\textbf{Book}}          \\ \midrule
\multicolumn{2}{c|}{\textbf{Models}}                                                                                          & AUC             & UAUC            & AUC             & UAUC            \\ \midrule
\multicolumn{1}{c|}{\multirow{4}{*}{\begin{tabular}[c]{@{}c@{}}\textbf{Model Structure}\\ \textbf{Variants}\end{tabular}}} & SeLLa-w/o  & 0.7357          & 0.7179          & 0.8391          & 0.6957          \\
\multicolumn{1}{c|}{}                                                                                    & SeLLa-Proj  & 0.7511          & 0.7333          & 0.8424          & 0.7225          \\
\multicolumn{1}{c|}{}                                                                                    & SeLLa-Warm & 0.7521          & 0.7371          & 0.8764          & 0.7444          \\
\multicolumn{1}{c|}{}                                                                                    & SeLLa-UI   & 0.7305          & 0.7076          & 0.8417          & 0.7186          \\ \midrule
\multicolumn{1}{c|}{\multirow{2}{*}{\begin{tabular}[c]{@{}c@{}}\textbf{Tunning Method}\\ \textbf{Variants}\end{tabular}}}  & SeLLa-W-UI & 0.7576          & 0.7458          & 0.8625          & 0.7350          \\
\multicolumn{1}{c|}{}                                                                                    & SeLLa-UI-W & 0.7539          & 0.7410          & 0.8797          & 0.7403          \\ \midrule
% \multicolumn{1}{c|}{\textbf{Special Variant}}                                                                     & LLM-emb   & 0.7365          & 0.7172          & 0.8357          & 0.6998          \\ \midrule
\multicolumn{1}{c|}{}                                                                                & SeLLa-Rec  & \textbf{0.7606} & \textbf{0.7464} & \textbf{0.8837} & \textbf{0.7459} \\ 
\bottomrule
\end{tabular}
\end{small}
\end{threeparttable}
}
\end{table}

\begin{figure}[htbp]
    \centering

    \begin{subfigure}[t]{0.45\textwidth}
        \centering
        \includegraphics[width=0.49\textwidth]{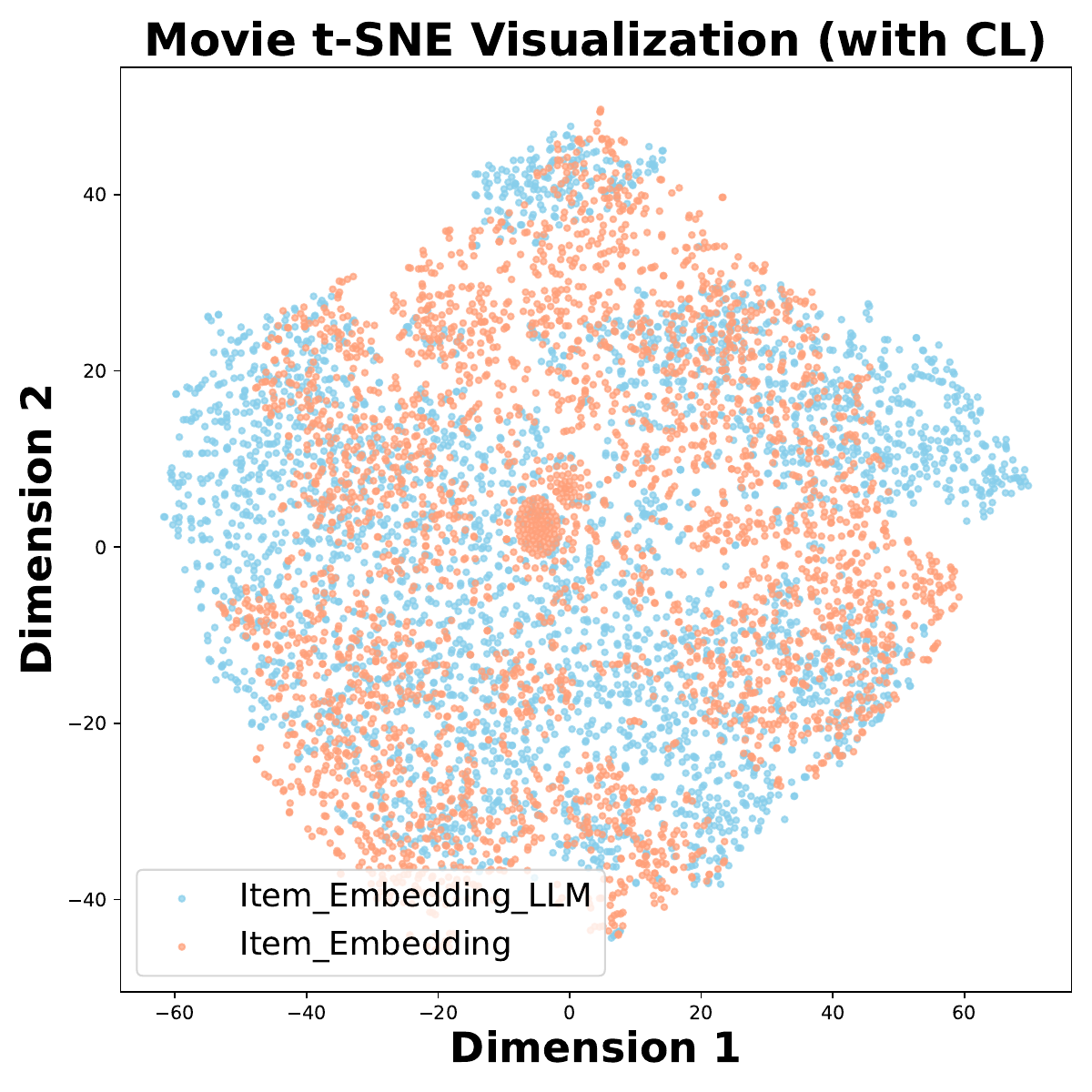} 
        \includegraphics[width=0.49\textwidth]{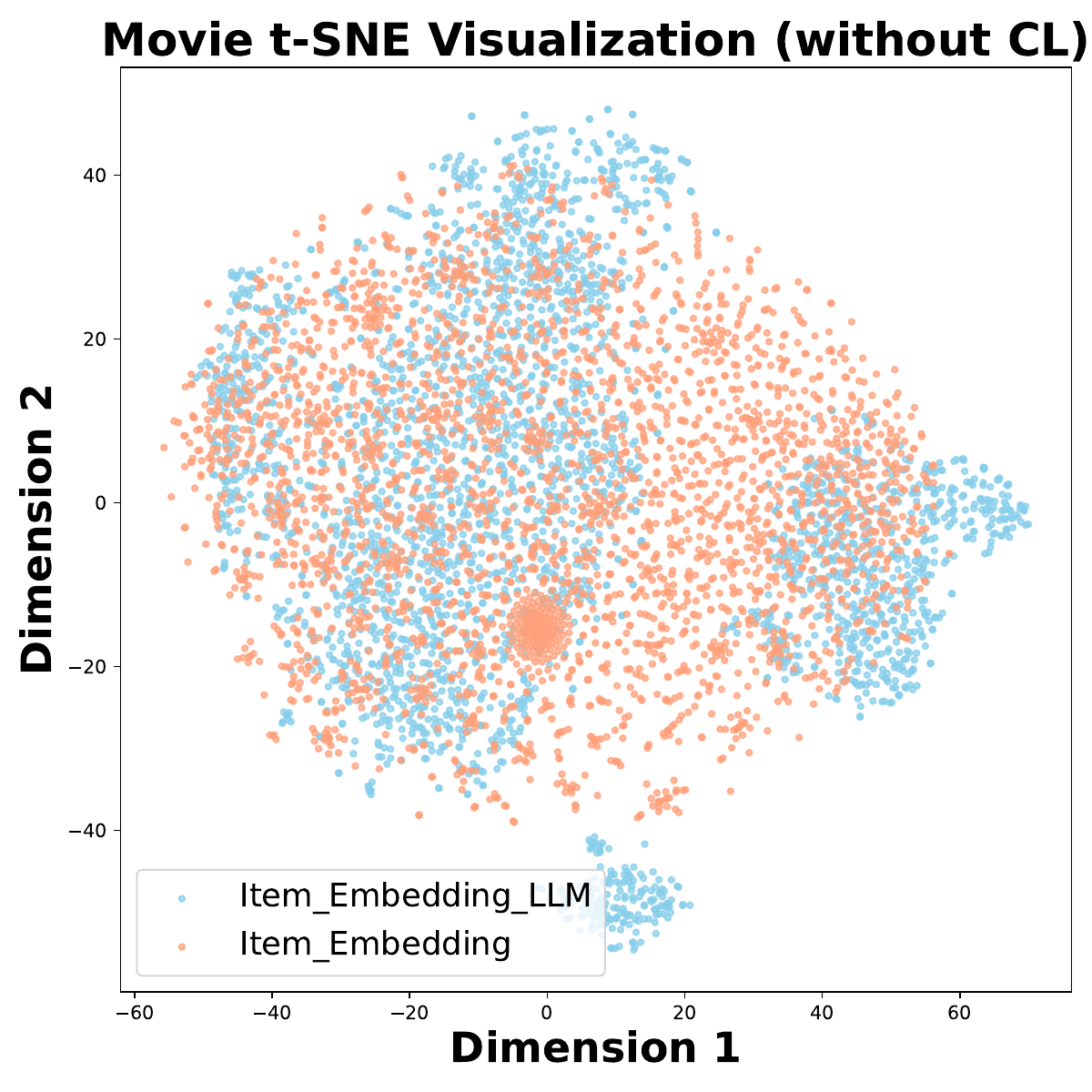} 
        % \caption{Distribution of two item embeddings with/without CL (Movie)} % 子标题
    \end{subfigure}
    \hfill

    \begin{subfigure}[t]{0.45\textwidth}
        \centering
        \includegraphics[width=0.49\textwidth]{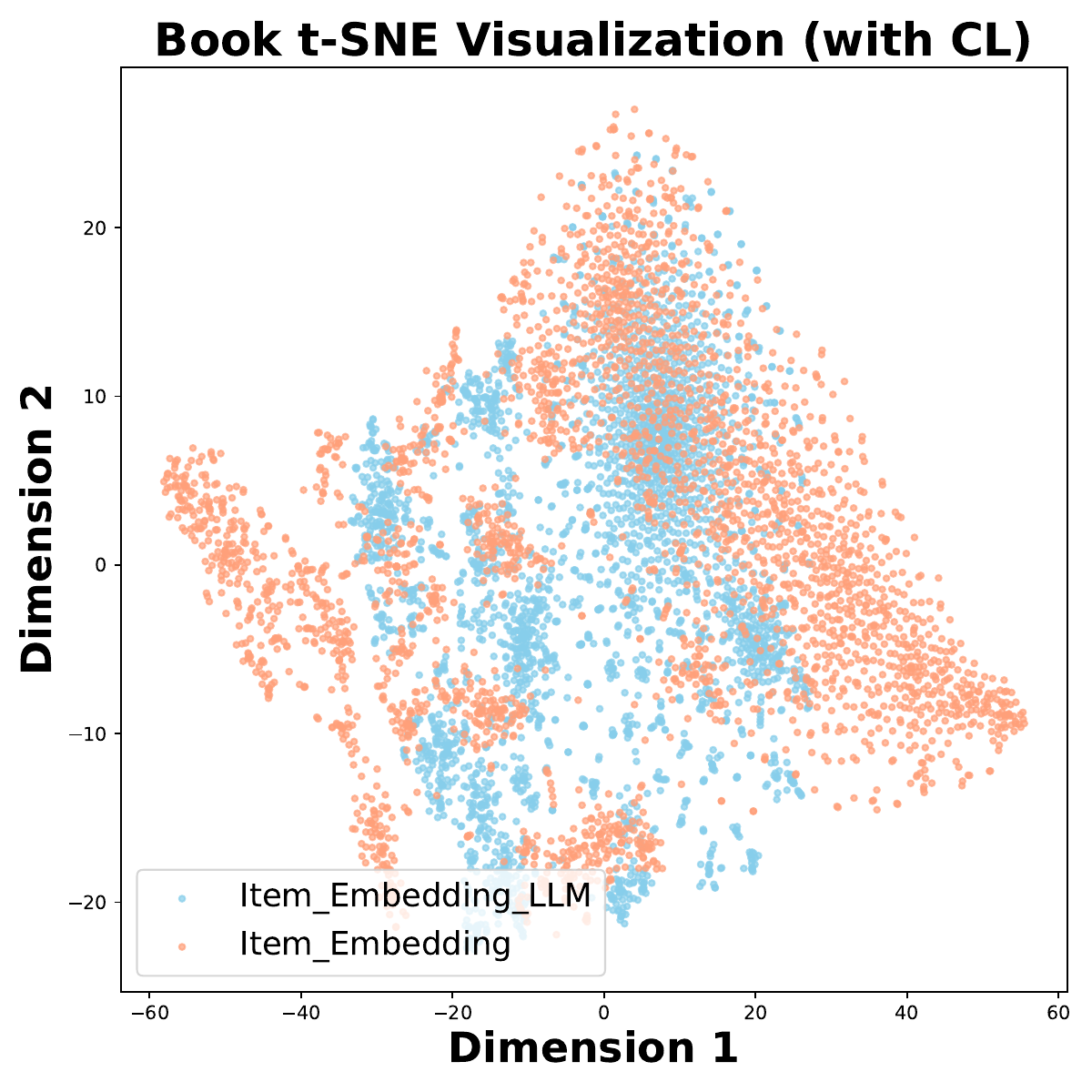} 
        \includegraphics[width=0.49\textwidth]{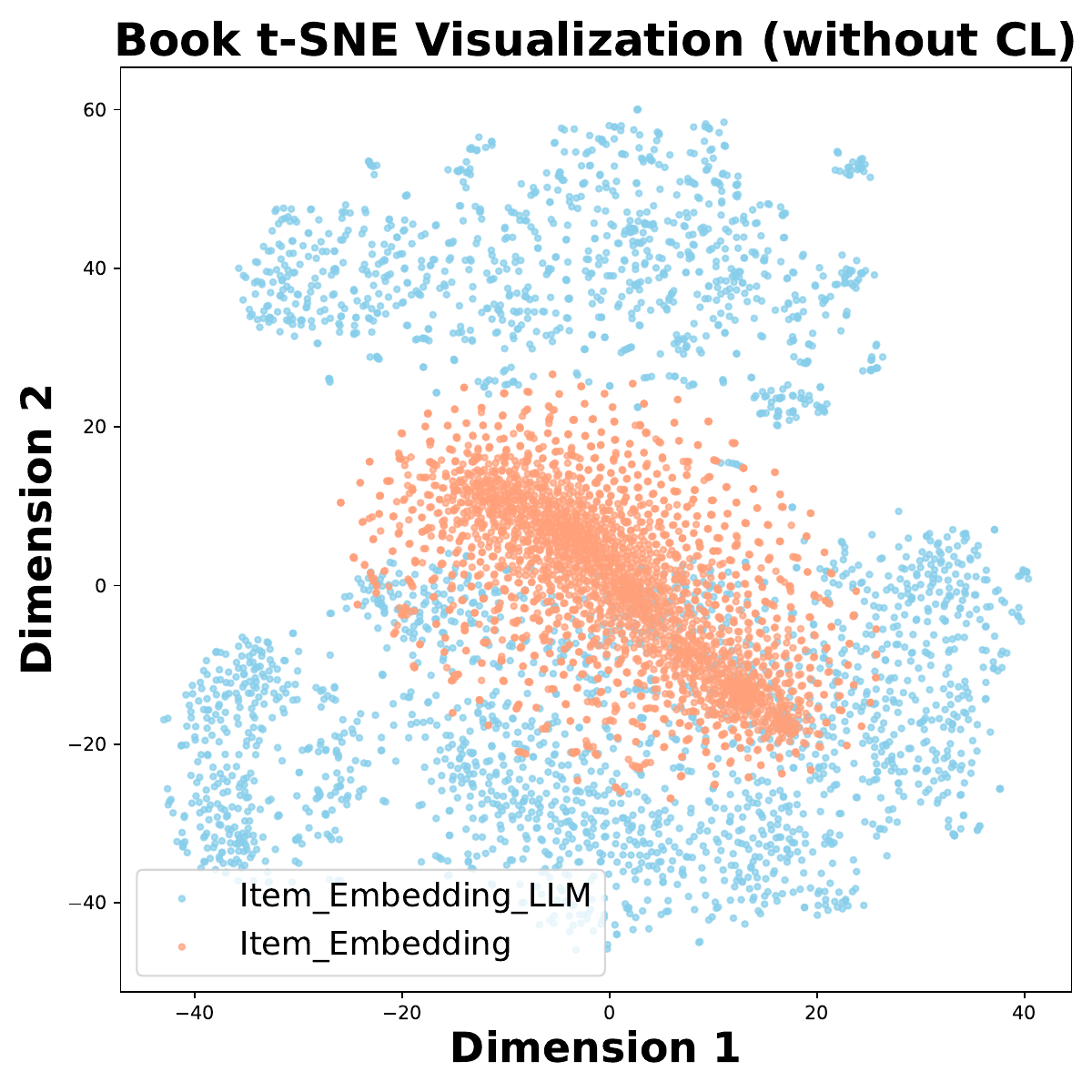} 
        % \caption{Distribution of two item embeddings with/without CL (Book)} % 子标题
    \end{subfigure}

    \caption{Distribution of two item embeddings in t-SNE}
    \label{fig:t-sne}
\end{figure}

\begin{figure}[htbp]
    \centering
    \begin{subfigure}[b]{0.23\textwidth}
        \centering
        \includegraphics[width=\textwidth]{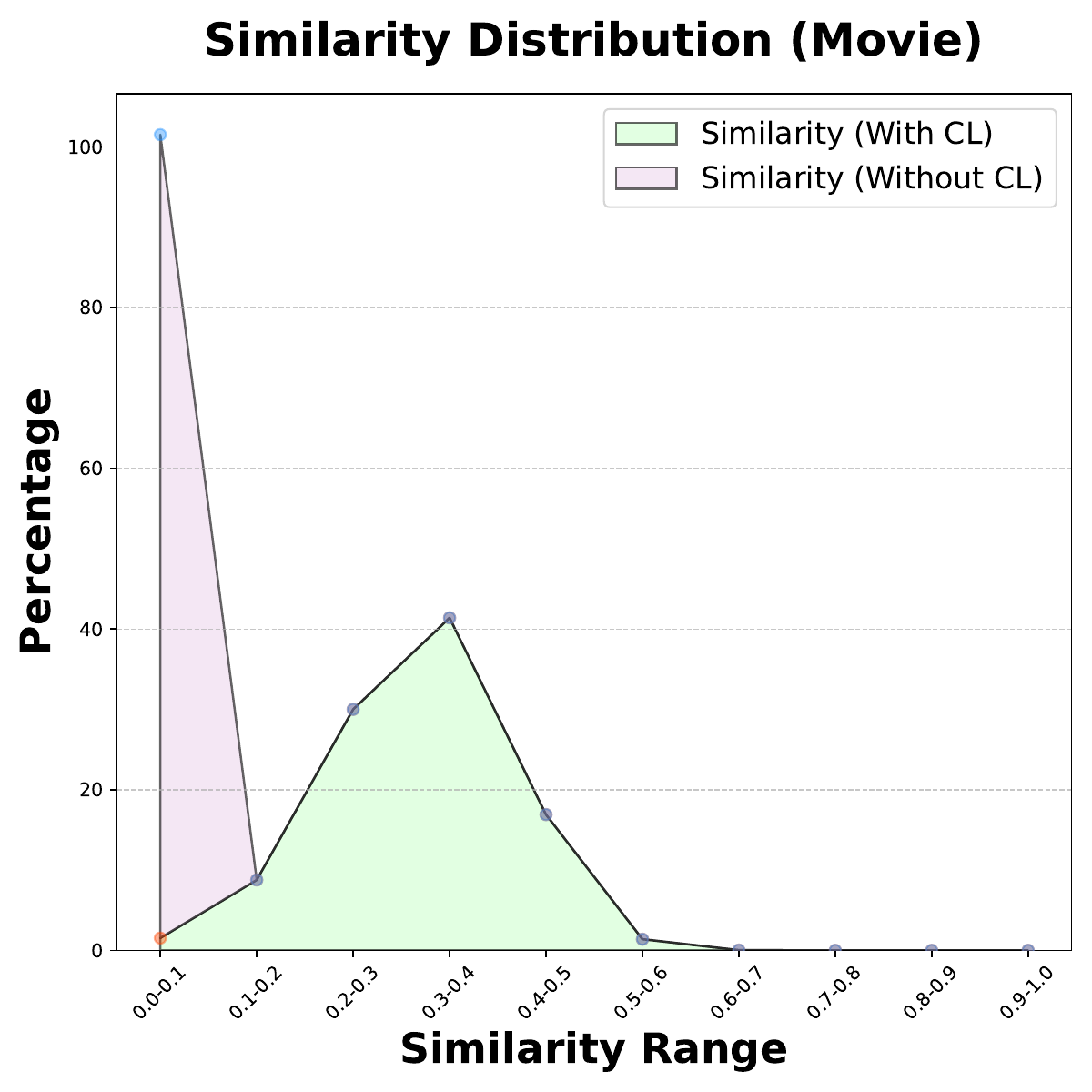} 
    \end{subfigure}
    \hfill
    \begin{subfigure}[b]{0.23\textwidth}
        \centering
        \includegraphics[width=\textwidth]{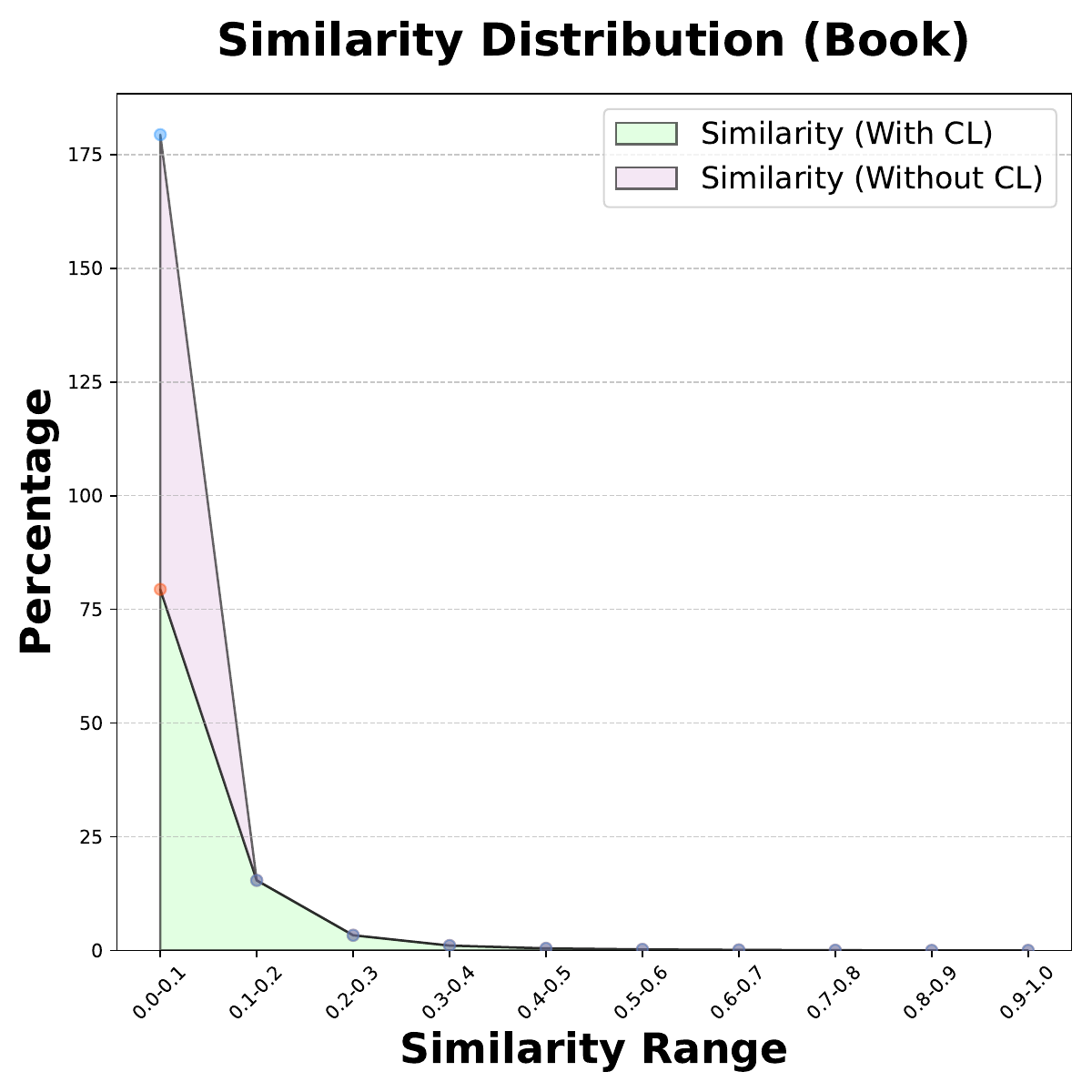} 
    \end{subfigure}
    
    \caption{Stacked area chart comparison for cos-similarity between two item embeddings.} 
    \label{fig:overall_cos}
\end{figure} 
\begin{figure*}[htbp]
    \centering
    \begin{subfigure}[b]{0.48\textwidth}
        \centering
        \includegraphics[width=\textwidth]{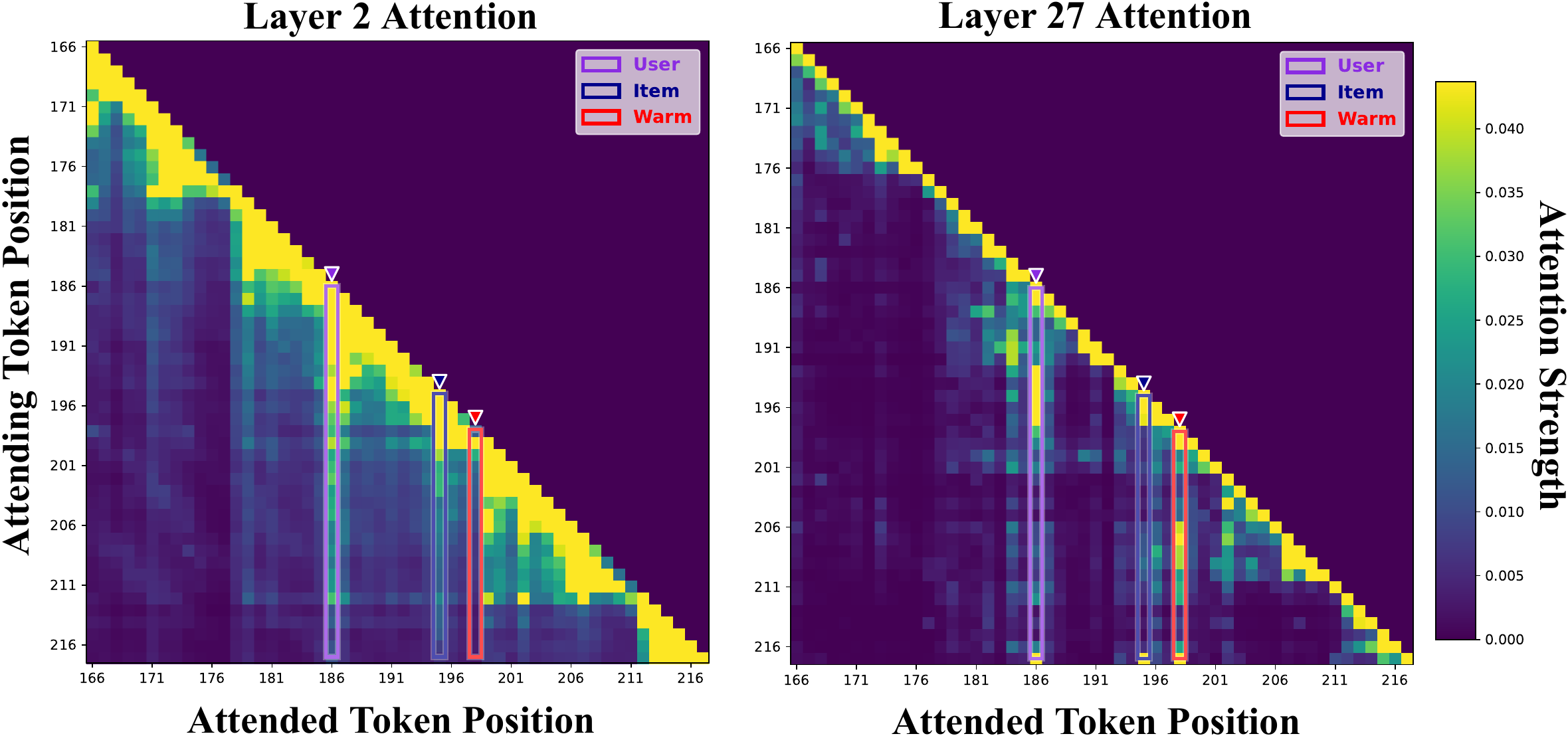}
        \caption{Attention Maps for User 288, Item 299 (Movie)}
        \label{fig:attn_movie}
    \end{subfigure}
    \hfill
    \begin{subfigure}[b]{0.48\textwidth}
        \centering
        \includegraphics[width=\textwidth]{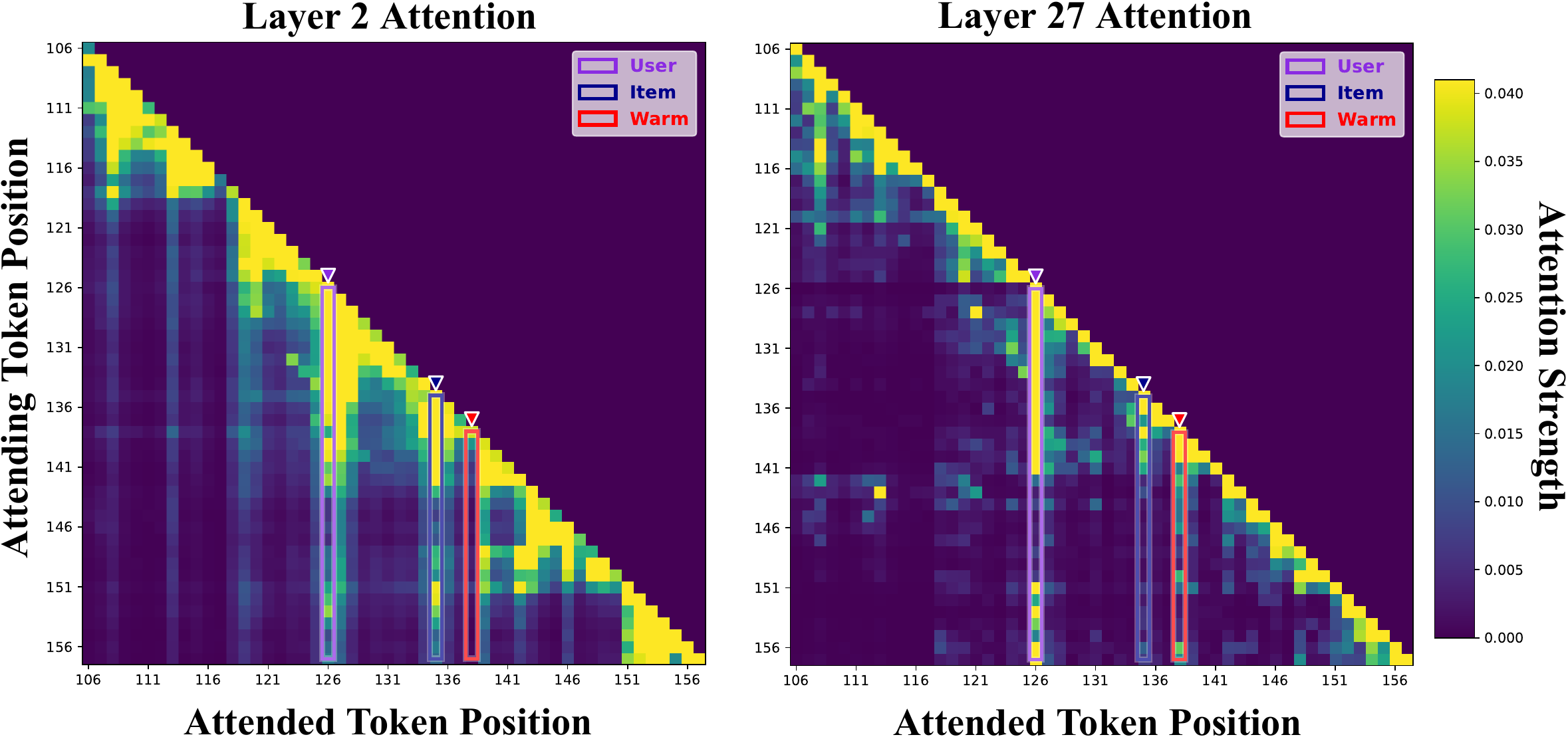}
        \caption{Attention Maps for User 3554, Item 422 (Book)}
        \label{fig:attn_book}
    \end{subfigure}
    % \vspace{-1em}
    \caption{Visualization of attention strength, achieved by averaging the attention weights across all heads at each layer. }
    \label{fig:attention}
\end{figure*}
\subsection{Ablation Study}
To evaluate the effectiveness of each setting in SeLLa-Rec, we design six primary variants.  
Four primary variants stem from modifications to the model architecture, while the remaining two are based on changes to the training methodology. 
The primary performance improvements of SeLLa-Rec stem from two key configurations: the pre-aligned training scheme for the collaborative model (Collab.)  and the introduction of three special tokens ($<User\_ID>, <Item\_ID>, <Warm\_ID>$) provided to LLMs. 
For these two configurations, we further propose two variants for each to investigate their impact. 
First, regarding the Collab.'s alignment training scheme, one variant involves removing alignment between Collab. and LLM altogether. 
While this appears to simplify the model by omitting a component, it effectively disables the subsequent alignment operations. 
Consequently, this variant essentially eliminates all changes brought by SeLLa-Rec and is referred to as \texttt{SeLLa-w/o}, which is the same as CoLLM. 
The second variant named \texttt{SeLLa-Proj} involves replacing the pre-warmed projection layer for Collab. with a randomly initialized projection layer. 
For the configuration involving the three special tokens, we directly evaluate the model's performance after removing them. 
Since there is an intrinsic dependency between the $<User_{ID}>$ and $<Item_{ID}>$, we treat them as a group. 
As a result, we define two additional variants: \texttt{SeLLa-Warm}, which corresponds to the model after removing the $<Warm\_ID>$, and \texttt{SeLLa-UI}, which corresponds to the model after removing the $<User\_ID>, <Item\_ID>$. 
The results in Tab.~\ref{tab:ablation} prove that any structural modifications negatively impact the model's performance, thereby validating the robustness and rationality of the SeLLa-Rec architecture. 

Essentially, the Warm-token and UI-token represent distinct types of knowledge, with differences in both their mapping layer structures and functions. 
This distinction raises the question of whether the current approach of training them simultaneously is indeed the most effective solution. 
Therefore, we compare the effectiveness of three different training strategies: the existing joint training scheme and two asynchronous training schemes (variants): 
(1) Warm-token-related components are trained first, followed by user-item (UI)-related components, which are referred to as \texttt{SeLLa-W-UI}. 
(2) UI-related components are trained first, followed by warm-token-related components referring to \texttt{SeLLa-UI-W}.
The experimental results demonstrate that our joint training method consistently performs best.

\subsection{Alignment Analysis}
During the downstream Collab. training, internal contrastive learning (CL) plays a crucial role in pre-aligning collaborative knowledge with LLM knowledge. 
To visualize this alignment effect, we employ t-SNE to examine the distributional differences between embeddings distilled from LLM and Collab.'s embeddings for items across two scenarios: with and without CL, as Fig.~\ref{fig:t-sne}. 
The results demonstrate that CL significantly narrows the distributional gap between the two embeddings. 
To assess these distributional differences quantitatively, we utilize stacked area charts in Fig.~\ref{fig:overall_cos} to illustrate the cosine similarity distribution between the two embeddings. 
The analysis reveals that after applying CL, the overall cosine similarity between the embeddings exhibits a notable increase, indicating enhanced alignment between the two representations. 

To verify whether better alignment can improve performance, we propose the following hypothesis: if aligned collaborative knowledge is closer to the semantic space of the LLM, then directly projecting this knowledge into three special tokens at the start of the training stage 3 (Step 0) should lead to improved model performance. Specifically, we evaluate three model variants on warm data at Step 0, as these data represent the core knowledge utilized in collaborative training. 
(1) Full SeLLa-Rec (denoted as CL+Projection), incorporating all alignment operations, including contrastive learning and pre-trained projection layers
(2) SeLLa-Rec with contrastive learning but a randomly initialized projection layer for Collab. (denoted as CL).
(3) A model without any alignment operations (structurally identical to CoLLM, denoted as w/o) 
The results in Tab.~\ref{tab:align_comparison} show that both alignment mechanisms enhance model performance at the initial step, indicating improved knowledge transfer from the collaborative component to the LLM. 
These empirical results, together with the above distributional analysis, provide compelling evidence that SeLLa-Rec's focus on semantic alignment between the LLM and collaborative components is fundamentally sound. 
\begin{table}[htbp]
\caption{Performance comparison on Movie-Warm and Book-Warm datasets.}
\label{tab:align_comparison}
\centering
\resizebox{0.35\textwidth}{!}{ % 调整表格宽度
\begin{threeparttable}
\begin{tabular}{l|l|ccc}
    \toprule
    Dataset & Metric              & CL + Projec & CL     & w/o  \\
    \midrule
    Movie-Warm &AUC    & \textbf{0.7421}          & 0.7393 & 0.7283 \\
    Book-Warm &AUC      & \textbf{0.8450}          & 0.8319 & 0.8158 \\
    \bottomrule
\end{tabular}
\end{threeparttable}
}
\end{table}

\subsection{Warm-Token Analysis}
To investigate how WARM-TOKEN influences LLM's recommendation-making mechanism, we perform an analysis by visualizing attention patterns across former and latter transformer layers of the LLM in Fig.\ref{fig:t-sne}. 
The examination focuses on the second layer (layer 2) and the last layer (layer 27). 
While our initial choice is the first layer (layer 1), we observe that attention patterns at the layer typically exhibit minimal differentiation. 
In the visualization figure, attention scores are represented by color intensity, with brighter colors indicating higher attention weights. 
The results clearly demonstrate that $<User\_ID>$, $<Item\_ID>$, and $<Warm\_ID>$ all play active roles in the attention mechanism. 
Sometimes, the attention weight assigned to $<Warm\_ID>$ even surpasses that of $<Item\_ID>$. 
This empirical evidence strongly suggests that WARM-TOKEN not only actively participates in the recommendation process but also contributes significantly to the system's overall performance. 

\section{Conclusion}
Existing LLM-based recommendation models often overlook the challenge of aligning the distinct semantic distributions of LLM and collaborative knowledge. 
The misalignment limits their ability to fully integrate collaborative signals into the semantic space of LLMs, resulting in suboptimal performance. 
To address this issue, we propose SeLLa-Rec, a novel framework designed to bridge this gap through semantic alignment between LLMs and Collab., enabling more effective knowledge integration and enhanced recommendation quality. 
SeLLa-Rec employs a three-layer architecture, where collaborative information originates from the bottom layer, is transformed into three special tokens in the middle layer, and is incorporated into the task prompt of the top-layer LLM backbone to enhance recommendations. 
SeLLa-Rec implements a hierarchical training strategy that guides the LLM to effectively utilize these special tokens to make better recommendations. 
The effectiveness of SeLLa-Rec and its components has been rigorously validated through comprehensive experiments and analyses on two public datasets: MovieLens-1M and Amazon Book. 

\bibliographystyle{ACM-Reference-Format}
\bibliography{ref}
\end{document}